\documentclass[12pt,preprint]{aastex}

\usepackage{natbib}

\slugcomment{to be submitted to ApJ}

\shorttitle{Elliptical Instability in Rotating Bars}

\shortauthors{ Ou, Tohline, \& Motl }

\begin{document}


\title{{Further Evidence for an Elliptical Instability in
Rotating Fluid Bars and Ellipsoidal Stars}}



\author{Shangli Ou\altaffilmark{1,2}, Joel E. Tohline\altaffilmark{1}
   \& Patrick M. Motl\altaffilmark{1}}

\altaffiltext{1}{Department of Physics \& Astronomy, 
      Louisiana State University, Baton Rouge, LA  70803}
\altaffiltext{2}{High Performance Computing, Center for Computation and Technology,
Information Technology Services,
Louisiana State University, Baton Rouge, LA, 70803;ou@cct.lsu.edu}


\begin{abstract}

Using a three-dimensional nonlinear hydrodynamic code, we examine
the dynamical stability of more than twenty self-gravitating,
compressible, ellipsoidal fluid configurations that initially have
the same velocity structure as Riemann S-type ellipsoids.  Our focus
is on ``adjoint'' configurations, in which internal fluid motions
dominate over the collective spin of the ellipsoidal figure;
Dedekind-like configurations are among this group.  We find that,
although some models are stable and some are moderately unstable,
the majority are violently unstable toward the development of $m=1$,
$m=3$, and higher-order azimuthal distortions that destroy the
coherent, $m=2$ bar-like structure of the initial ellipsoidal
configuration on a dynamical time scale.

The parameter regime over which our models are found to be unstable
generally corresponds with the regime over which incompressible
Riemann S-type ellipsoids have been found to be susceptible to an
elliptical strain instability \citep{LL96}. We therefore suspect
that an elliptical instability is responsible for the destruction of
our compressible analogs of Riemann ellipsoids. The existence of the
elliptical instability raises concerns regarding the final fate of
neutron stars that encounter the secular bar-mode instability and
regarding the spectrum of gravitational waves that will be radiated
from such systems.


\end{abstract}


\section{Introduction}

In a previous paper \citep[][hereafter paper I]{OU06}, we introduced
a new self-consistent-field technique that is capable of
constructing three-dimensional (3D) models of incompressible Riemann
S-type ellipsoids and compressible triaxial configurations that
share the same velocity fields as those of Riemann S-type
ellipsoids. These compressible triaxial configurations represent
fairly good quasi-equilibrium states and can be used to examine the
dynamical stability of Riemann S-type ellipsoids in the nonlinear
regime. With this in mind, a subset of our compressible models have
been evolved in a 3D hydrodynamics code \citep{MTF02} for $20 - 40$
dynamical times to test their dynamical stability. In this paper, we
present results from these hydrodynamic simulations.

Before going into the details of our simulations, it is worthwhile
spending some time reviewing previous studies of the stability
properties of Riemann S-type ellipsoids. Chandrasekhar and Lebovitz
carried out a vigorous analysis of the stability of Riemann S-type
ellipsoids with respect to second and third order harmonics (these
results are summarized in \cite{Ch69}). Some configurations,
including both rigidly-rotating Jacobi and stationary Dedekind
sequences, were found to be subject to a pear-mode (azimuthal mode
number $m=3$) instability beyond a certain point while proceeding
away from the axisymmetric Maclaurin spheroid sequence. Part of the
Jacobi and Dedekind sequences were also found to be unstable to a
dumbbell-shaped (azimuthal mode number $m=4$) mode; this has
provided some theoretical foundation for the fission theory of the
formation of binary stars \citep{DT85,Lebovitz87,Tohline02}. These
studies have assumed that the evolution of a rotating gas cloud
toward the bifurcation points of the $m=3$ pear-mode or $m=4$
dumbbell-mode proceeds in a quasi-stationary manner and on a time
scale that is fairly long compared to the dynamical time scale.

However, more recent linear stability studies \citep{LL96, LS99}
have revealed a new elliptical strain instability in Riemann S-type
ellipsoids that is associated with the noncircular fluid streamlines
that describe the internal motion of these configurations. According
to these studies, a fairly large fraction of all rotating ellipsoids
appear to be susceptible to the development of this elliptical
instability. It has a growth rate that is slow compared to the
rotation rate but is nevertheless dynamical. The fastest growth
rates are found among the adjoint configurations, in which the
internal motion dominates over the collective pattern motion of the
rotating bar. This elliptical instability was first discovered and
studied by \cite{P86} and \cite{B86} in the context of
two-dimensional (2D), incompressible fluid flows.
\cite{G93},
\cite{LPK93}, and \cite{RG94} extended it to astrophysics to study
tidally distorted accretion disks. They found that the instability
is three-dimensional (depends on the vertical dimension) and
approximately incompressible, and that it might also give birth to
$m=1$ internal waves in the absence of viscosity. It is subsequent
to this work on accretion disks that \cite{LL96} conducted their
study of Riemann S-type ellipsoids, for which elliptical flows are
generic.

The discovery of an elliptical strain instability in rotating
ellipsoidal (bar-like) structures brings into question \citep{LL96}
whether the paradigm for evolution of a secularly unstable star
driven by gravitational-radiation reaction (GRR) forces is viable.
According to this paradigm \citep{Ch70, DL77, LS95}, such a star --
in the form of a Riemann S-type ellipsoid -- would evolve on a
secular time scale toward the Dedekind sequence (for which the
pattern frequency of the bar-mode $\omega=0$), at which point the
evolution would stop because the gravitational field no longer
exhibits a time-varying quadrupole moment.  The analysis of
\cite{LL96} shows that the parameter space occupied by the
elliptical instability encompasses most of the Dedekind sequence.
Because the elliptical instability develops on a dynamical time
scale, it might prevent a GRR-driven figure from proceeding further
to the Dedekind sequence, especially if the elliptical instability
develops into the nonlinear regime. Because the results presented by
\cite{LL96} are based on linear analysis, they were not sure what
the nonlinear outcome of the elliptical instability would be.

In a recent nonlinear study of the development of the secular
bar-mode instability by \cite{OTL04}, an initially uniformly
rotating neutron star was driven by GRR forces into a differentially
rotating bar-like configuration that had a very low pattern
frequency. This bar-like configuration appeared to be a compressible
analogue of an adjoint configuration among the Riemann S-type
ellipsoids. However, the coherent bar-like structure was destroyed
very quickly by the nonlinear development of an unexpected
high-order instability that generated ``turbulence'' throughout the
internal structure of the bar. \cite{OTL04} suggested that this
turbulent-like instability was related to the elliptical
instability, but no definite conclusion could be drawn because only
one model was studied.  It was impractical to pursue a full
investigation across the entire parameter space of possible
ellipsoidal models because, at the time, the generation of one
single equilibrium bar-like model was computationally expensive; the
construction of even one equilibrium ellipsoidal model required
following the GRR-driven evolution of an initially axisymmetric
model, which takes a very long time even with an artificially
enhanced GRR force \citep{OTL04, SK04}.

With the new self-consistent-field technique introduced in Paper I,
we have been able to quickly build a large number of nearly
incompressible, quasi-equilibrium triaxial models. These
compressible models contain the same internal velocity flow-fields
as incompressible Riemann ellipsoids.  And although they are not
exact steady-state configurations, these models provide reasonably
good initial states for a study of the elliptical instability
because the elliptical instability develops on a dynamical time
scale. In what follows, we present hydrodynamic simulations of a
variety of these nearly incompressible, rotating bar-like models to
test their dynamical stability. We find that a significant fraction
are unstable to what appears to be the elliptical instability. We
introduce our initial models and analysis method in \S2; we present
results from hydrodynamic evolutions for different configurations in
\S3; and in \S4, we conclude and discuss issues relevant to the
evolutionary path of secularly unstable neutron stars.

\section{Initial Models and Methods}

\subsection{Initial Models}
We follow the notation defined in Paper I, where $a$, $b$, and $c$
are the three principal semiaxes of an ellipsoid and $c$ is the
rotation axis. In a frame of reference whose origin coincides with
the center of the ellipsoid and that is rotating with the pattern
frequency of the ellipsoid, $\omega$, the velocity field of a
Riemann S-type ellipsoid takes the form,
\begin{equation}
\vec{v} = \lambda (ay/b, -bx/a, 0) \,, \label{vel}
\end{equation}
where $\lambda$ is a constant that determines the degree of internal
motion of the fluid. This velocity field is designed so that
velocity vectors are everywhere tangent to a set of self-similar
concentric ellipses. (In the uniform-density, incompressible Riemann
configurations, these concentric ellipses also identify
equipotential contours.)  When $|\omega| > |\lambda|$, a model is
said to be a ``direct'' model, in which the figure rotation
dominates over internal motions. Conversely, when $|\omega| <
|\lambda|$, the configuration represents an ``adjoint'' model, in
which the internal fluid motion dominates over the figure rotation.

In Paper I, we constructed a variety of compressible models that
have the same velocity field as is found in incompressible Riemann
S-type ellipsoids, that is, with the velocity field defined by
Eq.~(\ref{vel}). These models satisfied the steady-state Euler's
equation exactly, but only satisfied the steady-state continuity
equation approximately. However, as discussed in Paper I, it was
found that the deviation from the steady-state continuity equation
was small for models that obeyed a relatively stiff equation of
state (EOS).  For our present study, we have used the same
self-consistent-field technique described in Paper I to construct
more than forty quasi-equilibrium, compressible analogues of Riemann
S-type ellipsoids to serve as initial states for a series of
hydrodynamic simulations.  The models were constructed using two
different EOS ($n=0.5$ and $n=1$ polytropes), with a variety of
initial axis ratios -- Figure \ref{triaxial} displays the range of
selected ($b/a$, $c/a$) pairs -- and, for a given set of axes, both
``direct'' and ``adjoint'' configurations were constructed.  This
set of models was selected to cover a reasonably large fraction of
the parameter space that was investigated by \cite{LL96}, keeping in
mind that their linear analysis was confined to incompressible
figures.

The stability of roughly twenty-five of these initial models was
examined using a 3D hydrodynamic code, as explained below. We
focused on analyzing the stability of adjoint models with an $n=0.5$
polytropic EOS, but several direct models with an $n=0.5$ EOS and
one adjoint model with an $n=1$ EOS were included in the mix.  For
the sake of brevity, only the results of six model evolutions are
presented in detail in \S\ref{Sec:HydroResults}. Table
\ref{table:models} lists the key parameters for these six
representative models. In addition to the name that has been
assigned to each model and the values of five parameters that
already have been defined ($n$, $b/a$, $c/a$, $\omega$, $\lambda$),
Table \ref{table:models} specifies each initial model's ratio of
rotational to gravitational potential energy $T/|W|$, total angular
momentum $J_\mathrm{tot}$, mean density $\rho_\mathrm{mean}$ 
, and the ratio of the pattern
period $T_p\equiv 2\pi/\omega$ to the dynamical time
$T_\mathrm{dyn}\equiv [\pi G \rho_\mathrm{mean}]^{-1/2}$. The
measured growth rates $T_1$ and $T_3$ of unstable modes with
azimuthal mode numbers $m=1$ and $m=3$ are also tabulated (see \S3
for further elaboration). 
Dimensional quantities are in the hydro code unit
for which the gravitational constant $G$, the central density $\rho_c$,
and the cylindrical radius of the entire grids $\varpi_{grid}$ 
are all set to unity.
If the first letter in a model's name is
{\bf D}, the model is a ``direct'' configuration, whereas an {\bf A}
denotes an ``adjoint'' configuration.  Among this group of six
models, we note that only model {\bf A067} has a polytropic index
$n=1$; the other models all have $n= 0.5$.

\begin{deluxetable}{cccccccccccc}
\tablecolumns{12} \tablewidth{0pt} \tablecaption{Parameters of
selected initial models\label{table:models}} \tablehead{
\colhead{model} & \colhead{$n$} & \colhead{$b/a$} & \colhead{$c/a$}
& \colhead{$\omega$} & \colhead{$\lambda$} & \colhead{$T/|W|$} &
\colhead{$J_{\mathrm{tot}}$} & \colhead{$\rho_{\mathrm{mean}}$} &
\colhead{$T_p/T_\mathrm{dyn}$} & \colhead{$T_1/T_\mathrm{dyn}$} &
\colhead{$T_3/T_\mathrm{dyn}$} } \startdata

   {\bf D105} & 0.5 &0.59  &  0.487  & 0.911 &   0.114 &   0.105 &  0.00809 &  0.526  & 1.4 & - & -\\
   {\bf D080}  & 0.5&0.49  &  0.487  & 0.885 &   0.230 &   0.080 &  0.00497 &  0.529  & 1.5 &  - & - \\

   {\bf A010} & 0.5& 0.74  &  0.821 & -0.610  &-0.875  &0.010  &7.24e-3  &0.543   & 2.1 &  - &  -\\
   {\bf A067} & 1.0& 0.90  &  0.692 & -0.168  &-0.796  &0.067  &0.0087  &0.280  & 5.6  &  -  &  -\\

   {\bf A134}  & 0.5&0.74  &  0.487 & -8.1e-3& -0.911 & 0.134 & 0.0133  & 0.519  & 158 &  3.3 &  3.4\\

   {\bf A100}  & 0.5&0.59  &  0.487 & -0.114 & -0.911 & 0.100 & 7.21e-3  & 0.526  & 11&  1.4 &  1.3\\

\enddata
\end{deluxetable}


\subsection{Methods}

In order to examine their relative stability, each model was evolved
on a cylindrical coordinate mesh using the 3D, Newtonian,
finite-volume computational fluid dynamics technique described in
detail by \cite{MTF02}. In each simulation, the adopted grid
resolution was $66 \times 102 \times 128$ in the $\varpi$
(cylindrical radius), $z$ (vertical), and $\phi$ (azimuthal)
directions, respectively.

We monitor the time-evolution of the general structure of each model
by measuring the Fourier amplitude of various ``modes'' having
azimuthal quantum numbers $m$ in the following fashion: At each
instance of time, the azimuthal density distribution in a ring of
fixed $R$ and $z$ can be decomposed into a series of azimuthal
Fourier components via the relation,
\begin{equation}
  \rho(R,z, \phi) = \sum\limits_{m=-\infty}^{+\infty}
      C_m(R,z) e^{i m \phi} \, ,
\end{equation}
where the complex amplitudes $C_m$ are defined by the expression,
\begin{equation}
   C_m(R,z) = \frac{1}{2\pi} \int_0^{2\pi}
      \rho(R,z, \phi) e^{-i m \phi} d \phi \,.
\end{equation}
In our simulations, the time-dependent behavior of the magnitude of
this coefficient, $|C_m|$, is monitored at a variety of $(R,z)$
locations to measure the growth-rate of various modes. The
time-varying amplitudes shown in our figures, below, are averages of
$|C_m|$ over the entire volume.

Although this diagnostic tool permits us to study the behavior of
structure having a wide variety of $m$ values, in practice we have
focused our attention on the lowest order modes for this study.  For
example, a trace of the time-evolution of $|C_2|$ lets us monitor
the overall ellipticity of each model.  The $m=2$ mode should
dominate initially because of each model's initial bar-like
structure.  If this amplitude remains at a fairly constant level in
a given simulation, we conclude that the configuration is stable;
otherwise, it is unstable. In practice, we have found that when the
$m=2$ amplitude drops significantly in a model, that model
simultaneously displays growth of other azimuthal modes.

In their analysis of the elliptical instability, \cite{LL96}
generally found that the unstable mode with the fastest growth rate
was the $m=3$ mode.  In their study, the $m=1$ mode was not
discussed because, in incompressible configurations, an $m=1$ mode
normally refers to a translation of the center of mass of the
configuration. However, in the study of an elliptical instability in
tidally distorted accretion disks by \cite{G93} and \cite{LPK93}, it
was found that an eccentric $m=1$ mode may be important. Further
investigation shows that the instability has some dependence on the
third (vertical) dimension, because it disappeared in 2D simulations
\citep{RG94}. With these results in mind, we have focused our
attention on the development of the $m=1$ and $3$ modes in our 3D
ellipsoidal model evolutions.

\section{Results From Hydrodynamic Simulations}\label{Sec:HydroResults}

We present our results from hydrodynamic evolutions for direct and
adjoint configurations in the next two subsections, respectively.

\subsection{Results For Direct Configurations}
\begin{figure}[ht]\epsscale{1} \plotone{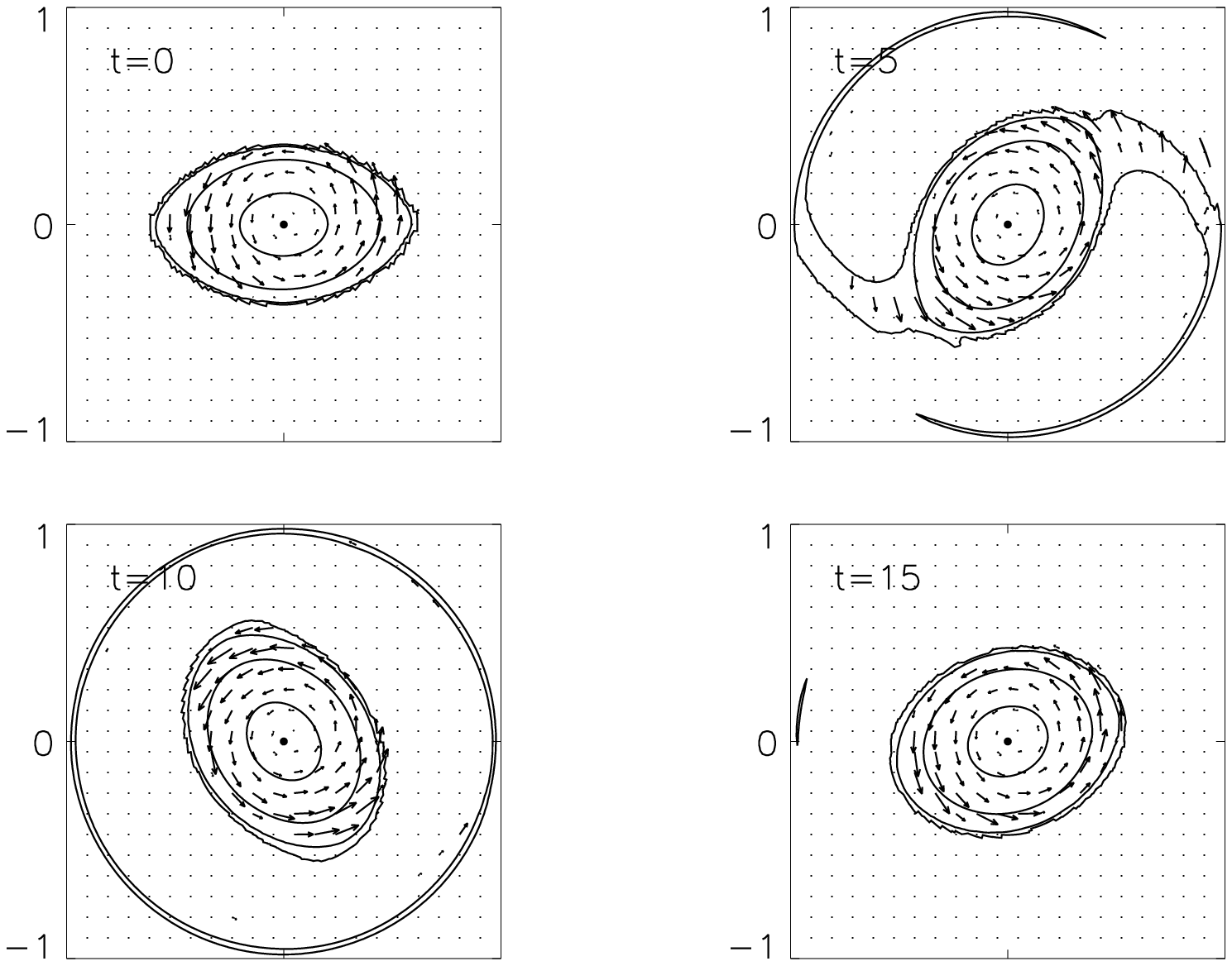}
\caption{ Equatorial-plane iso-density contours and velocity fields
    in the inertial frame for model {\bf D105}. Density contours correspond to
    $\rho/\rho_\mathrm{max} = 0.01, 0.1, 0.5$, and $0.8$ from the outermost
    to the innermost shell, and time is given in units of $T_\mathrm{dyn}$.
    The pattern rotation of the overall ellipsoidal configuration and the
    internal fluid motion are both prograde (counterclockwise) in this
    ``direct'' model configuration.  An accompanying MPEG movie
    shows the entire evolution of this model, through $t \approx 24
    ~T_\mathrm{dyn}$; 3D, rather than 2D, iso-density contours are shown in the movie
    without the accompanying flow field.
    \label{den:d105}}
\end{figure}

\begin{figure}[ht]\epsscale{1} \plotone{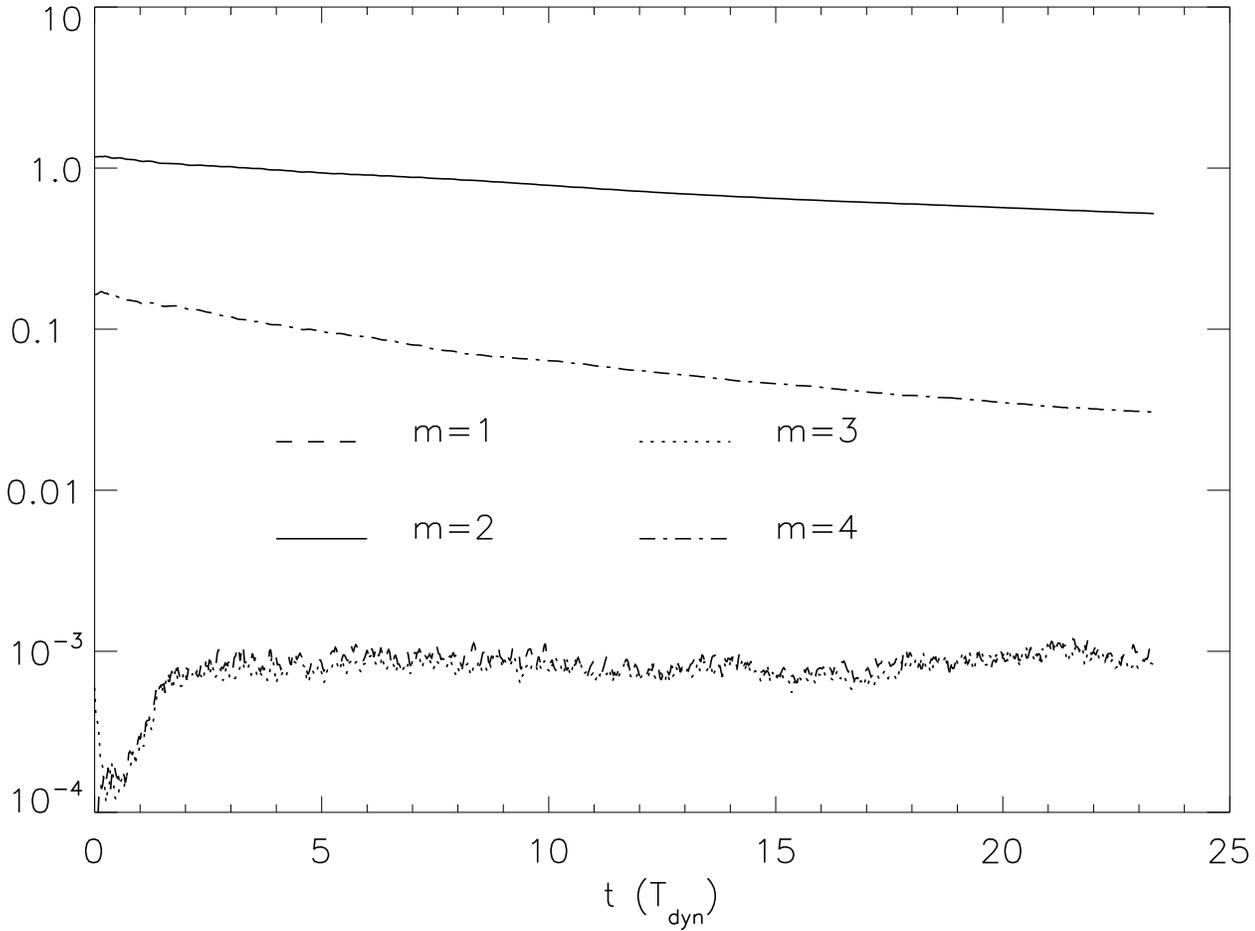}
\caption{
    Time-evolution of the amplitudes $|C_m|$ of the $ m=1,2,3,$ and 4 modes for
    model {\bf D105}. The amplitudes of the $m=2$ and 4 modes decay gradually;
    the $m=1$ and 3 modes remain at a fairly constant, low-amplitude level,
    which suggests that this ``direct'' model is stable against the elliptical instability.
    \label{mode:d105} }
\end{figure}

In this subsection, we present results from two direct
configurations, models {\bf D105} and {\bf D080}, which were evolved
for about 24 $T_\mathrm{dyn}$ and 17 $T_\mathrm{dyn}$, respectively.
Figure \ref{den:d105} shows snapshots in time of equatorial-plane
iso-density contours and the inertial-frame velocity field for model
{\bf D105}. 
(In the online version of this paper, the figure caption
includes a link to an MPEG movie showing the entire evolution of
this model.) 
This model is stable; the coherent bar-like structure
and elliptical flows are well maintained throughout the 24
$T_\mathrm{dyn}$ followed by our simulation. The time-varying
Fourier mode amplitudes displayed in Figure \ref{mode:d105} show
that the $m=2$ bar-mode remains the dominant mode throughout this
evolution, with an amplitude that remains approximately three orders
of magnitude higher than those of the $m=1$ and $3$ modes. Although
the $m=4$ mode has a larger amplitude than the odd modes, the fact
that its behavior follows that of the $m=2$ mode suggests that it is
a harmonic of the $m=2$ mode and that there is no independent $m=4$
mode.

Although we have categorized model {\bf D105} as stable, a slow
decay of the bar-mode amplitude is evident in Figure 2. Because
other modes, including the $\mathrm{m}=1$ and $3$ modes, do not grow
throughout the evolution, this slow decay of the bar-mode probably
results from the small violation of the steady-state continuity
equation in our initial model configuration.  (This, in turn,
results from the fact that the 3D iso-density surfaces in our
compressible models deviate from self-similar ellipses or homeoidal
shells; see Paper I for a more detailed discussion.) Small
violations in the surface layers cause very low density material to
be shed outward from both ends of the object and form two low
density spiral waves (see the panel labeled $t=5$ in Figure 1).
Because this material carries relatively high specific angular
momentum, the overall system loses angular momentum, but only on a
very long time scale. We suspect that this is the major cause of the
gradual decay of the $m=2$ bar-mode in this model evolution.

\begin{figure}[ht]\epsscale{1} \plotone{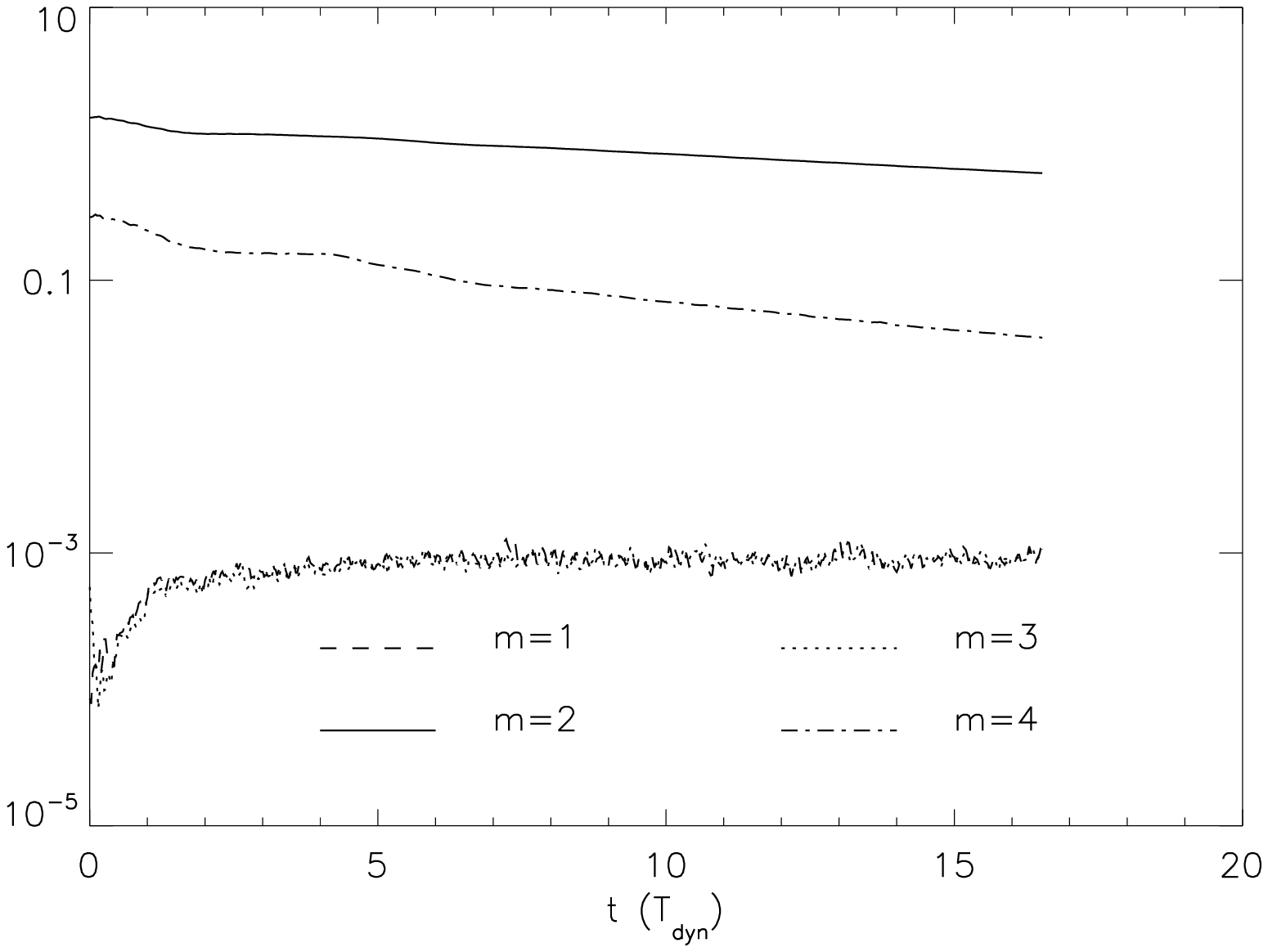}
\caption{
    The same as Figure \ref{mode:d105}, but for model {\bf D080}.
    This ``direct'' model also appears to be stable against the elliptical
    instability.
    \label{mode:d080}}
\end{figure}

Figure \ref{mode:d080} shows a similar Fourier mode analysis for
model {\bf D080}. Again, the ${m}=2$ bar-mode dominates the entire
evolution and shows a slow decay; amplitudes of the odd azimuthal
modes remain at roughly constant ``noise'' levels.

The results from the evolution of these two direct configurations
suggest that, although our initial model configurations do not
satisfy the steady-state continuity equation exactly, the models are
still fairly good quasi-equilibrium states. In both cases, the
coherent bar-like structures were well maintained through $\sim 20$
$T_\mathrm{dyn}$. Hence, we are confident that initial models
generated by the technique described in Paper I can be used to study
the dynamical stability of compressible Riemann S-type ellipsoids.
Furthermore, the fact that the elliptical instability sets in on a
dynamical time scale will allow us to use these models to detect the
elliptical instability, if it exists.

According to \citet[see their Figure 2 for a summary of their
analysis of direct configurations]{LL96}, the elliptical instability
only arises in a small portion of the parameter space that is
occupied by direct configurations, and its growth rate is relatively
slow there. This makes it relatively difficult for our current study
to identify the exact parameter-space boundaries between stable and
unstable direct configurations.
Furthermore, the method described in Paper I can only be used to
construct compressible models having axis ratios $b/a \gtrsim 0.5$,
whereas most of the domain that we expect to be effected by the
elliptical instability in direct configurations has $b/a < 0.5$.
This also makes direct configurations unsuitable for identifying the
existence of the elliptical instability.

Therefore, we have chosen to focus our study on an analysis of
adjoint configurations where the domain of the instability is much
wider, according to \citet[see especially their Figure 3a, which is
reproduced here as Figure \ref{LLplot}]{LL96}. More importantly, the
growth rate of the instability is expected to be faster in adjoint
configurations. This helps us to conduct such a survey with a
limited evolution time that is allowed by our current computational
resources. In the next subsection, we present simulations of adjoint
configurations, which indeed reveal the development of
turbulence-like flows that we suspect are associated with the
elliptical instability.

\subsection{Results of Adjoint Configurations and Existence of a
Turbulent-like Instability}

In this subsection, we present results from 3D hydrodynamic
evolutions of the four adjoint configurations listed in Table
\ref{table:models}. In light of previous studies \citep{G93,LL96},
it is expected that the $m=1$ and $3$ modes will be the fastest
growing modes, if an elliptical instability sets in. Therefore,
rapid growth of these two modes can serve as a compass pointing to
the elliptical instability. With this in mind, in the analysis below
we will categorize our results based on the behavior of these two
modes. Based on the final outcome and behavior of the $m=1$ and $3$
modes, evolutions of our initially quasi-equilibrium compressible
adjoint configurations can be divided into three categories: stable
models, moderately unstable models, and violently unstable models.

\begin{figure}[ht]\epsscale{1} \plotone{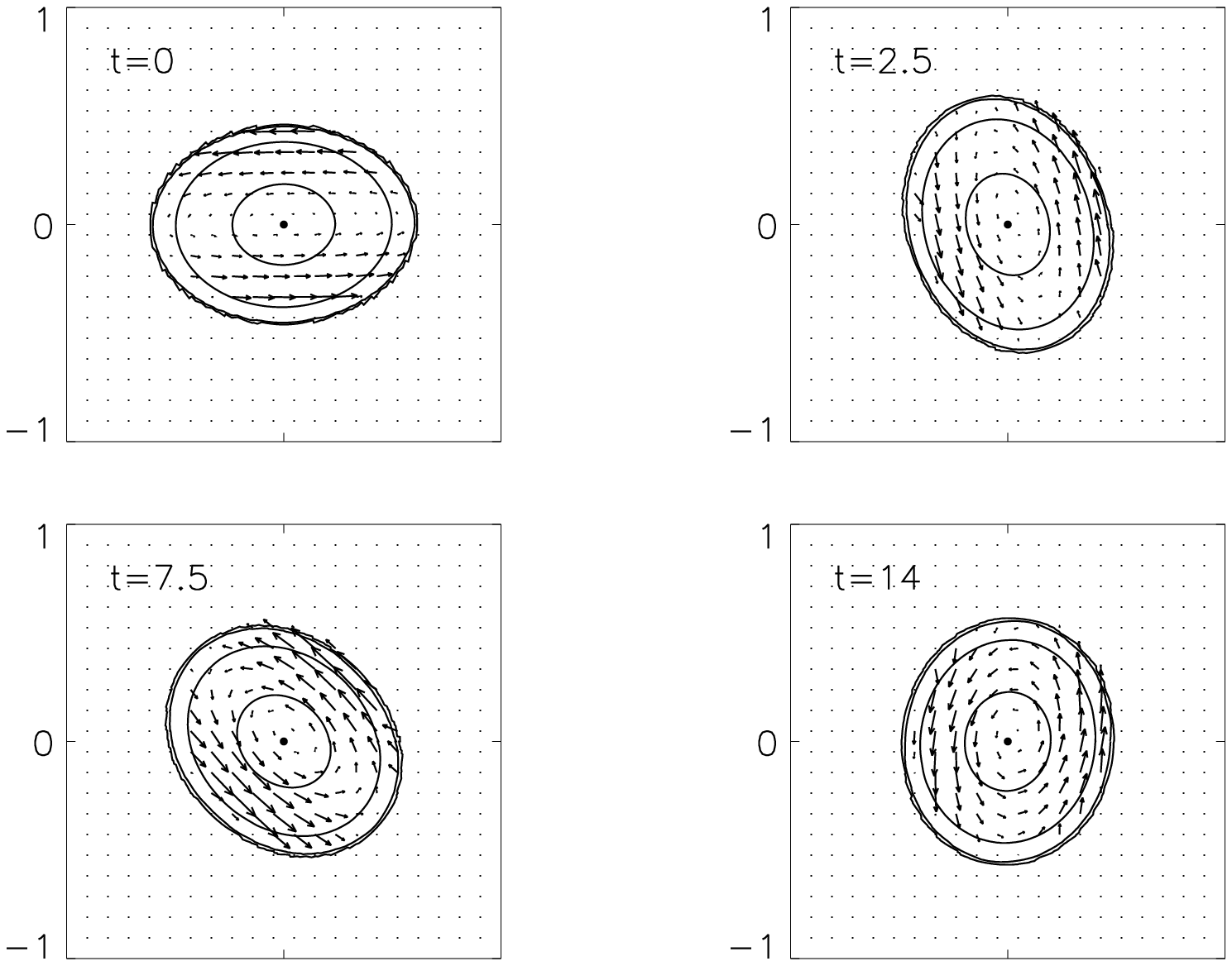}
\caption{
    The same as Figure \ref{den:d105}, but for model {\bf A010}.
    In this ``adjoint'' model, the internal fluid motion is prograde
    (counterclockwise), whereas the overall motion of
    the ellipsoidal configuration is retrograde.
    An accompanying MPEG movie shows the entire evolution of this
    model, through $t\approx 26~T_\mathrm{dyn}$. 
    \label{den:a010}}
\end{figure}

\begin{figure}[ht]\epsscale{1} \plotone{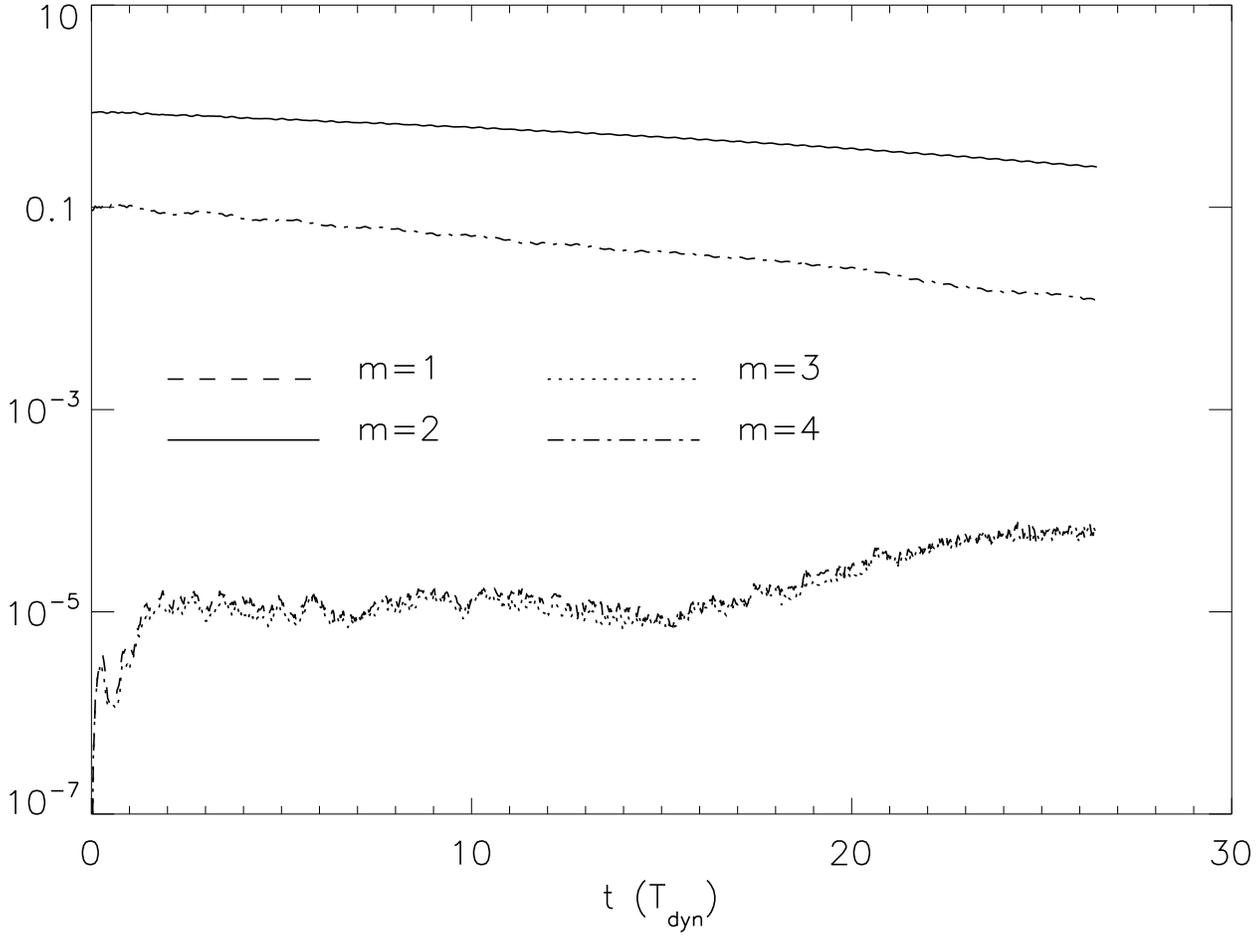}
\caption{
    The same as Figure \ref{mode:d105}, but for model {\bf A010}.
    All the modes appear to be stable in this ``adjoint'' model.
    \label{mode:a010} }
\end{figure}

\begin{figure}[ht]\epsscale{1} \plotone{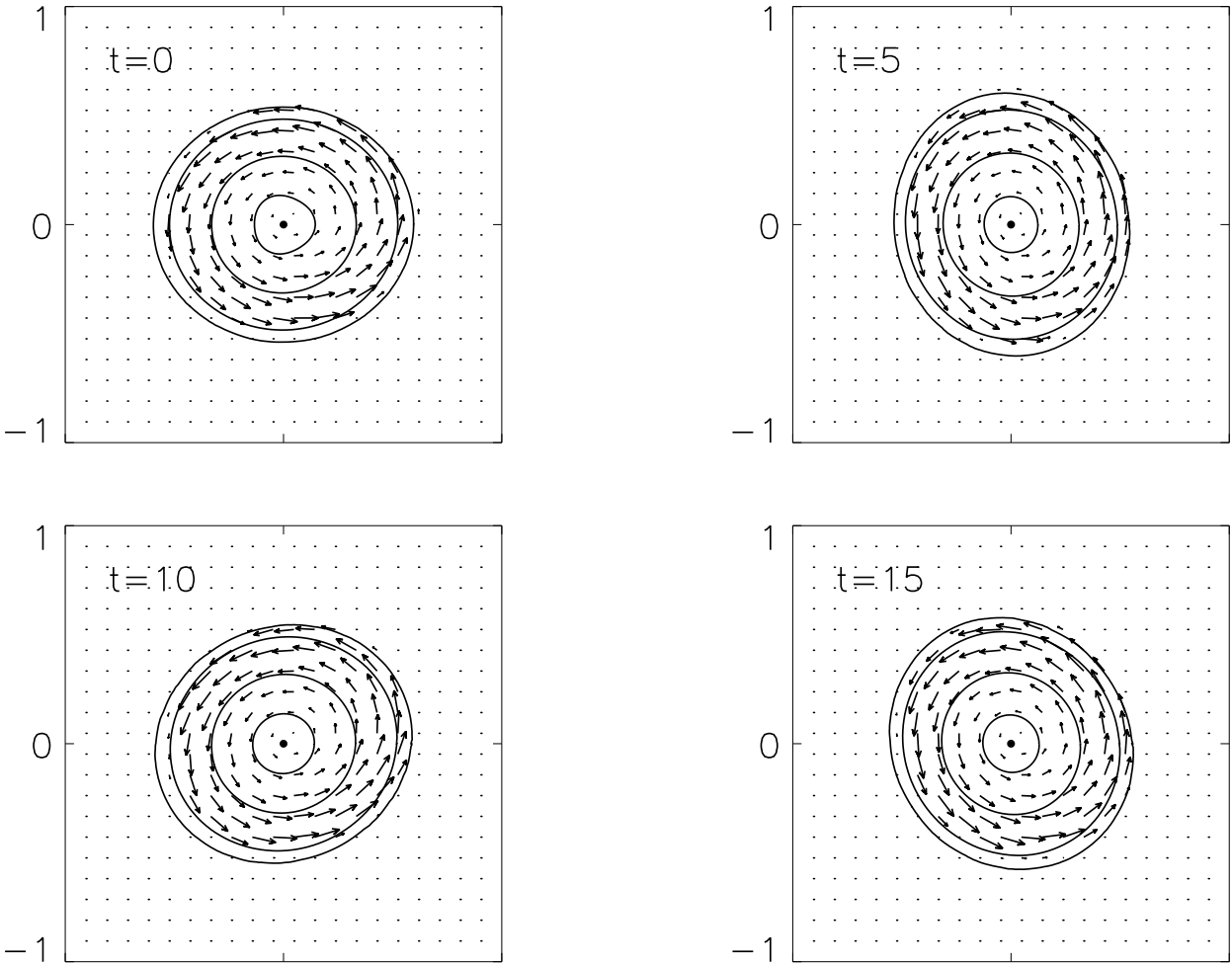}
\caption{
    The same as Figure \ref{den:d105}, but for model {\bf A067}.
    In this ``adjoint'' configuration, the fluid is moving prograde
    (counterclockwise) while the overall ellipsoidal pattern is
    spinning retrograde.
    \label{den:a067}}
\end{figure}

\begin{figure}[ht]\epsscale{1} \plotone{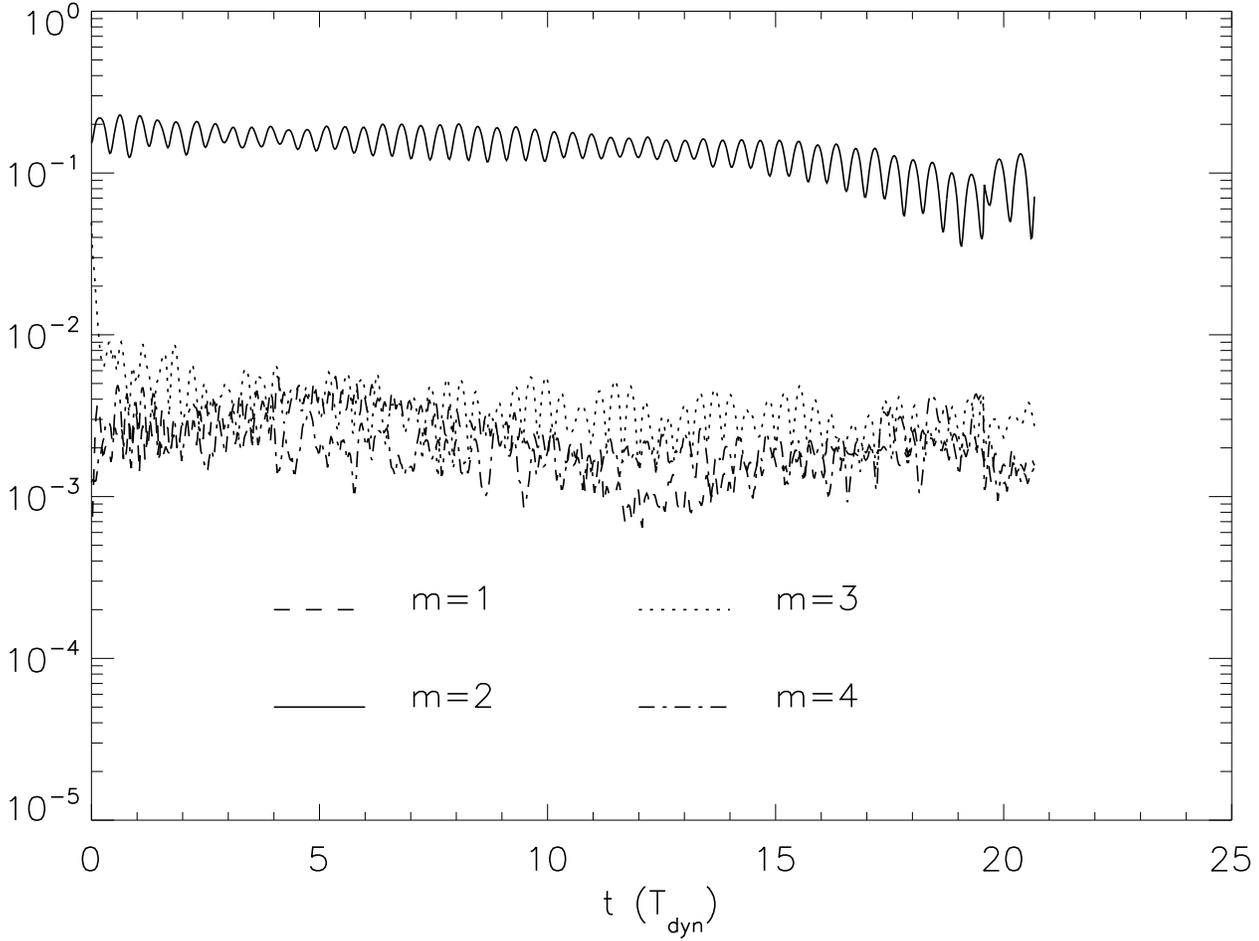}
\caption{
    The same as Figure \ref{mode:d105}, but for model {\bf A067}.
    All the modes appear to be stable.
    Note that a low-amplitude, $m=3$ perturbation was imposed on this
    model at time $t=0.$
    \label{mode:a067} }
\end{figure}

The first category --- stable models --- consists of evolutions in
which the internal elliptical flow remains stable. One evolution in
this family is model {\bf A010}, with $b/a=0.74$. Note from the
information in Table \ref{table:models} that this model has a
somewhat prolate structure with $c>b$. Figure \ref{den:a010} shows
2D, equatorial-plane iso-density contours of this model at four
different times. 
(The figure caption in the online version 
includes a link to an MPEG movie showing the entire evolution of
this model.) 
The ellipsoidal shape is well preserved, at least
up through the end of our simulation $\sim 26 ~T_\mathrm{dyn}$, and
the internal elliptical flow pattern remains stable. As suggested by
the Fourier mode analysis shown in Figure \ref{mode:a010}, the $m=2$
bar-mode dominates throughout the entire evolution. There is no
indication that the $m=1$ and $3$ modes are growing; their
amplitudes remain fairly constant at ``noise'' level throughout the
evolution. Figures \ref{den:a067} and \ref{mode:a067} show similar
plots for another stable model, namely, model {\bf A067} with
$b/a=0.90$. This model was perturbed with an initial low-amplitude
$m=3$ perturbation, but its bar-like structure remains stable during
the evolution. We note that the bar-mode appears to undergo a
low-amplitude oscillation.
As we have discussed in Paper I, initially small deviations from the
steady-state continuity equation can generate surface waves in
adjoint models. The oscillations observed here in the amplitude of
the $m=2$ bar-mode is indeed a reflection of these low-amplitude
surface waves traveling along the elliptical surface. The initial
violation of the steady-state continuity equation is larger in
models with a softer EOS (see Paper I).  It is therefore not
surprising that oscillations of the bar-mode amplitude in model {\bf
A067} are somewhat larger than in other models because model {\bf
A067} was constructed using a relatively soft ($n=1$ polytropic)
EOS.

\begin{figure}[ht]\epsscale{1} \plotone{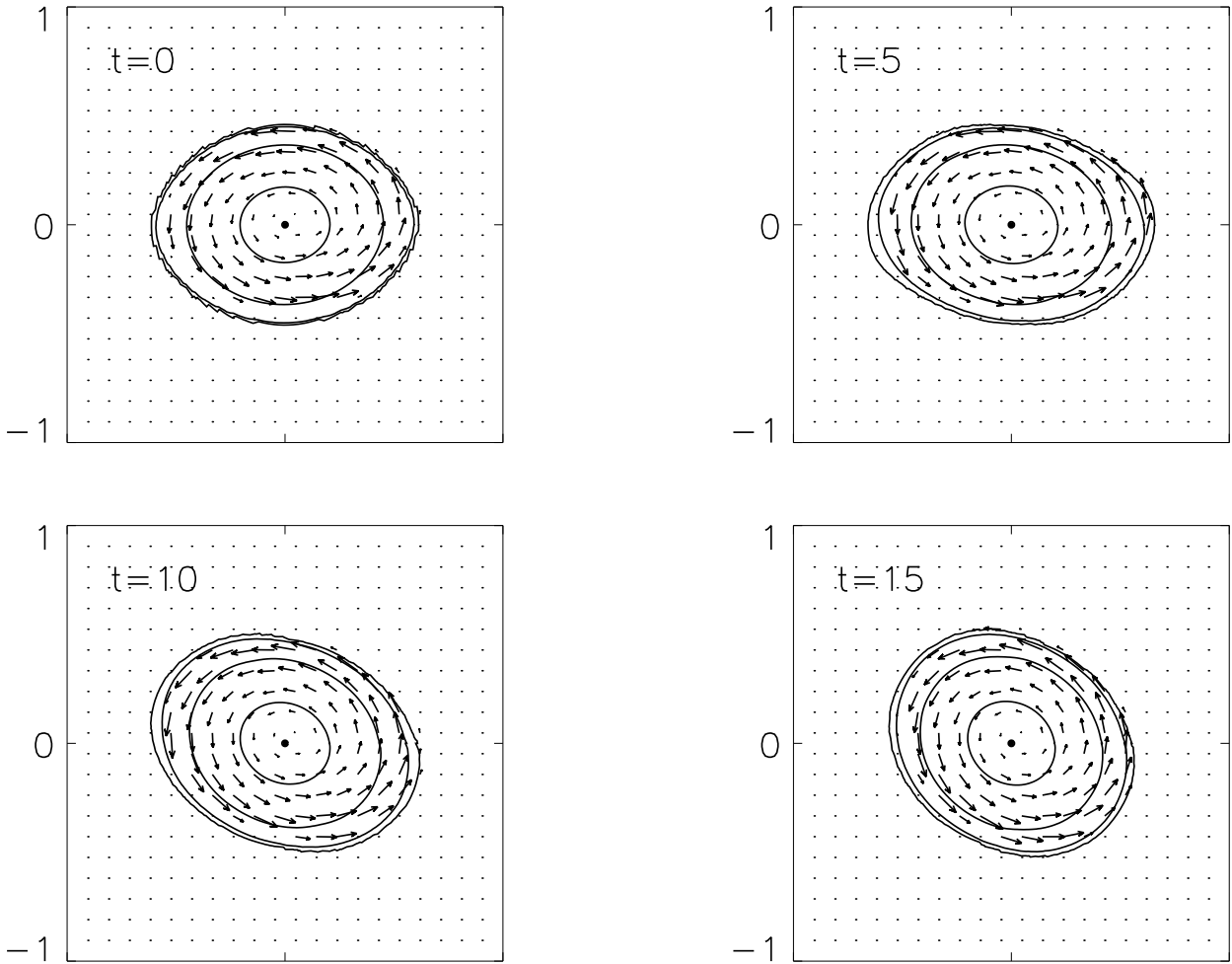}
\caption{
    The same as Figure \ref{den:d105}, but for model {\bf A134}.
    The internal fluid motion is prograde (counterclockwise), but
    this ``Dedekind-like'' model has an almost vanishing angular
    pattern frequency, $\omega=-0.0081$; as seen here, the figure
    pattern shifts clockwise by only $\approx \pi/4$ radians in
    the first $15~T_\mathrm{dyn}$.   From the four snapshots displayed
    here, the bar structure seems to be very stable,
    but from the mode analysis shown in Figure \ref{mode:a134},
    we conclude that the model is moderately unstable.  An
    accompanying MPEG movie shows the entire evolution of this
    model, through $t\approx 40 ~T_\mathrm{dyn}$.
    \label{den:a134}}
\end{figure}

\begin{figure}[ht]\epsscale{1} \plotone{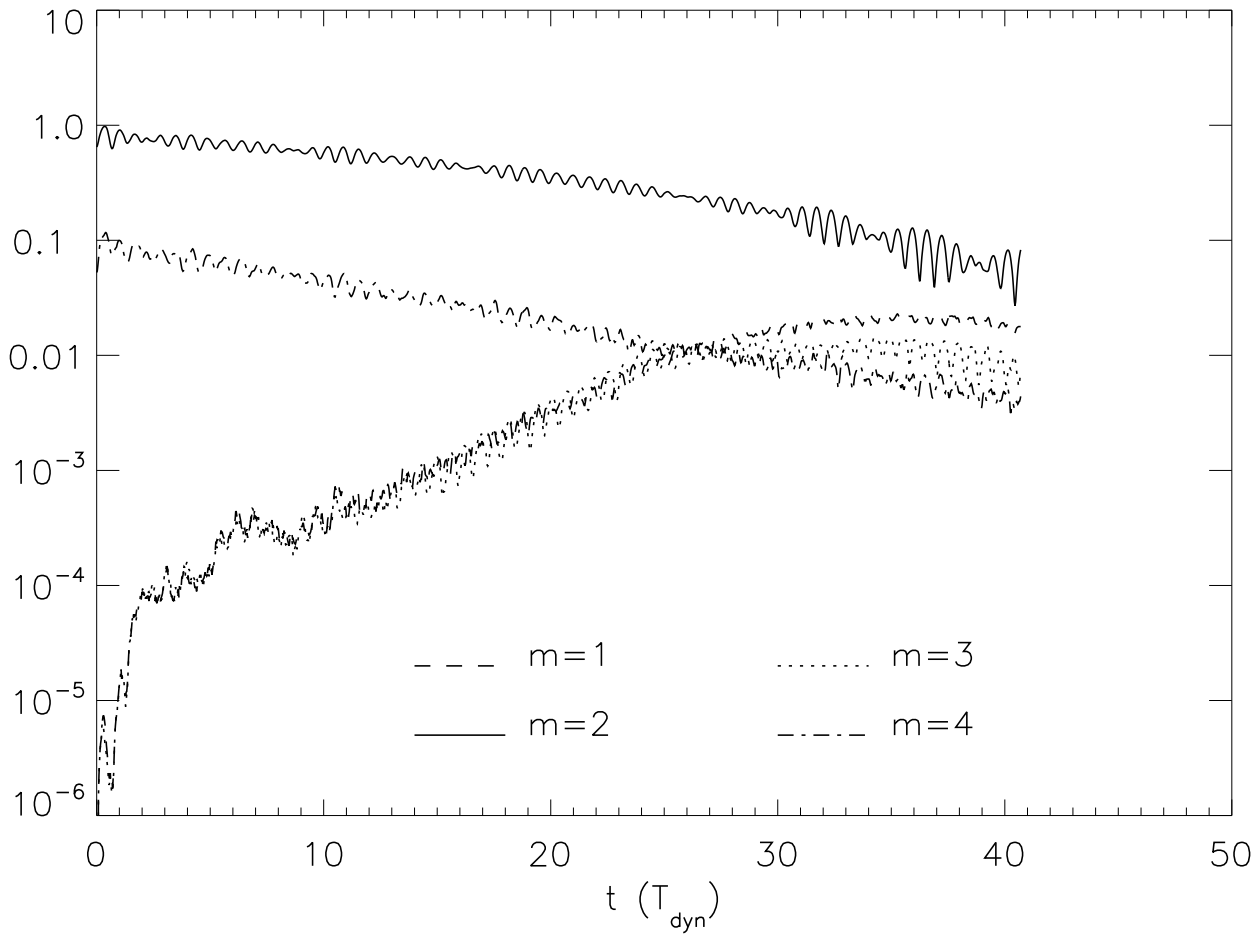}
\caption{
    The same as Figure \ref{mode:d105}, but for model {\bf A134}.
    The amplitude of the $m=2$ mode remains fairly constant throughout
    most of this evolution.  The $m=1$ and 3 modes also seem to be stable
    at early times, but ultimately they both exhibit a period of exponential
    growth (with a relatively slow growth time),
    which suggests that the model is moderately unstable.
    \label{mode:a134}}
\end{figure}

The second category of adjoint evolutions consists of configurations
that are moderately unstable. In these models, the elliptical flow
appears to remain stable as judged by the behavior of the $m=2$
Fourier mode.  However, the $m=1$ and $3$ modes are observed to grow
on a long time scale. Model {\bf A134} with $b/a=0.74$ belongs to
this family. Figure \ref{den:a134} shows 2D iso-density contours of
this model in the equatorial plane at different evolutionary times.
(An MPEG movie showing part of the evolution of this model
is available in the figure caption of the online version.) 
We note that this model has an almost vanishingly small bar pattern
frequency, $\omega=-0.0081$, so it is very nearly a Dedekind-like
object; as seen in Figure \ref{den:a134}, the bar-like pattern
shifts retrograde (clockwise) through only $\approx \pi/4$ radians
over $15~T_\mathrm{dyn}$! Through this initial $\sim15 ~
T_\mathrm{dyn}$, the coherent ellipsoidal structure and elliptical
flow pattern remain stable. However, as revealed by the Fourier mode
analysis in Figure \ref{mode:a134}, at later times the ${m}=1$ and
$3$ modes both exhibit exponential growth, albeit at a very slow
growth rate. We note that the amplitudes of these odd modes appear
to be tied to each other; even their individual growth rates are
almost the same. (As recorded in Table \ref{table:models}, the
e-folding time for these odd modes are $T_1= 3.3 T_\mathrm{dyn}$,
and $T_3= 3.4 T_\mathrm{dyn}$.)  This suggests that the mechanism
that is responsible for driving these modes is probably the same.
Their amplitudes do not exceed that of the $m=2$ mode at the end of
this simulation.
Hence, we have identified this model as being moderately unstable.
Across the parameter space of our investigation (see, for example,
Figure \ref{triaxial}), models belonging to this category can be
regarded as a transition from stable to violently unstable models.

\begin{figure}[ht]\epsscale{1} \plotone{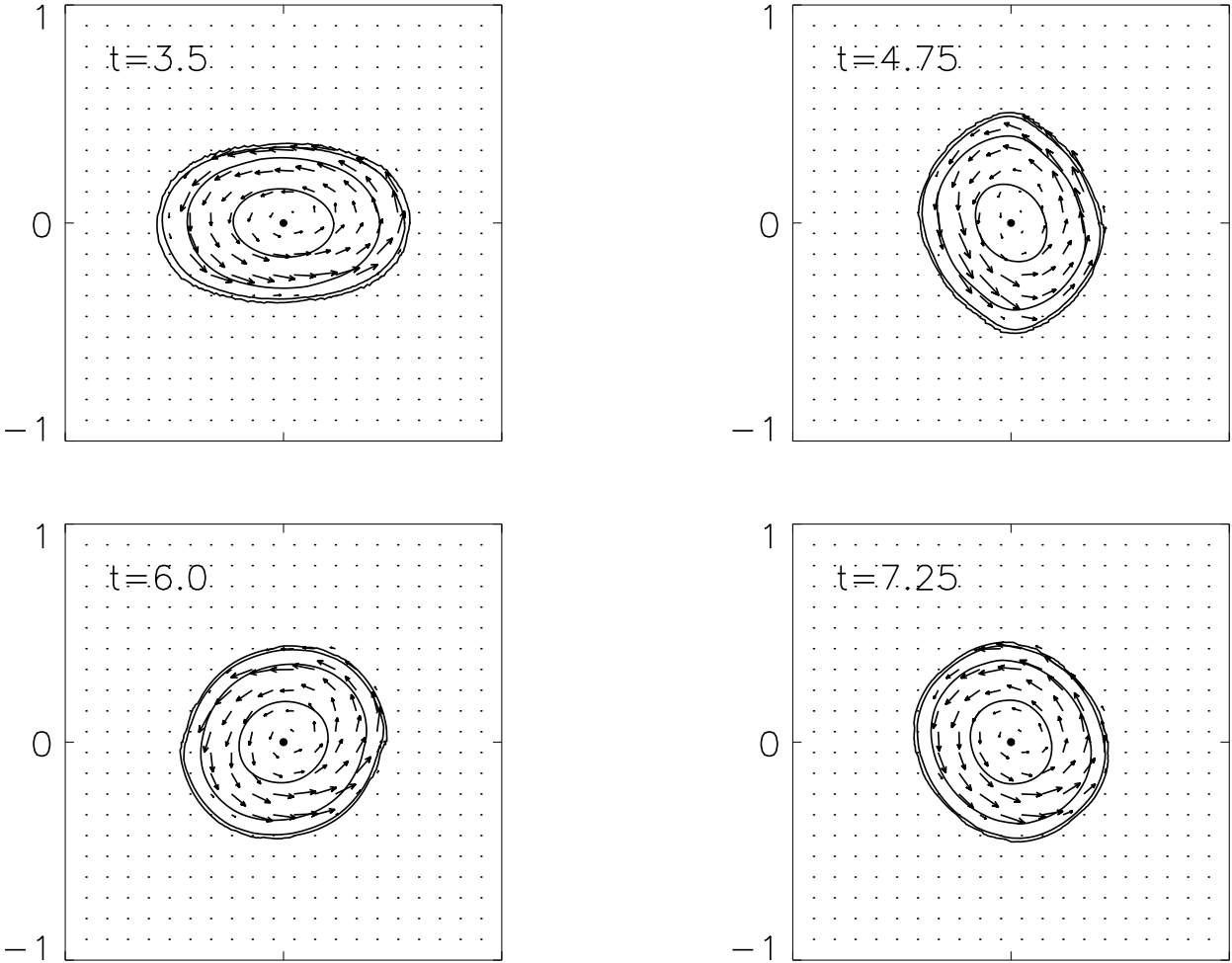}
\caption{
    The same as Figure \ref{den:d105}, but for model {\bf A100}.
    These frames have been selected between $ 3 < t/T_\mathrm{dyn} < 8$,
    during which the initial bar-like structure is entirely destroyed
    by some turbulent instability.  An accompanying MPEG movie shows
    the entire evolution of this model, through
    $t\approx 50 ~T_\mathrm{dyn}$. 
    \label{den:a100}}
\end{figure}

\begin{figure}[ht]\epsscale{1} \plotone{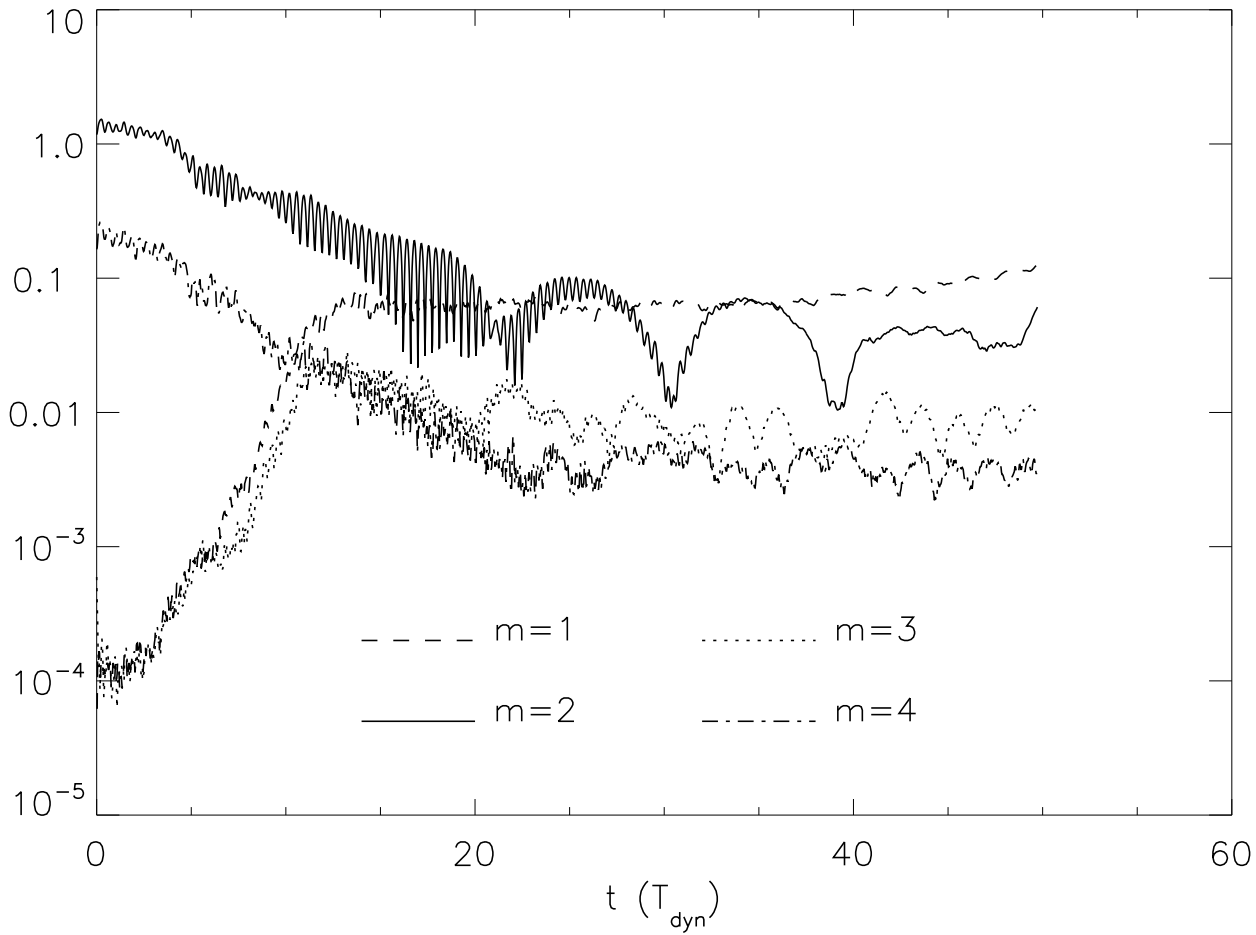}
\caption{
    The same as Figure \ref{mode:d105}, but for model {\bf A100}.
    The $m=1$ and 3 modes grow very quickly, in concert with the
    overall development of turbulent motions inside the
    configuration.  At the same time, the $m=2$ mode also decays
    very quickly.  The initial bar-like structure is destroyed
    in less than 8 $T_\mathrm{dyn}$.
    \label{mode:a100} }
\end{figure}

The third category consists of models that are violently unstable to
an instability that leads to apparently turbulent flow throughout
the configuration. The $m=2$ bar-mode is quickly destroyed in a few
dynamical times; simultaneously, both the $m=1$ and $3$ modes
exhibit extremely fast exponential growth.  One of the most unstable
models in our investigation is {\bf A100} with $b/a=0.59$. The
time-evolution of equatorial-plane iso-density contours of this
model is shown in Figure \ref{den:a100}. 
(The entire evolution is shown in a MPEG movie in the figure caption
of this plot in the online version of this paper.)
The initial ellipsoidal
structure exists for only $\sim 3 T_\mathrm{dyn}$, then it is
suddenly destroyed by some type of ``turbulent'' instability. After
the bar-like structure is destroyed, the whole configuration settles
into an entirely turbulent state, exhibiting small-scale vortices
and eddies. The Fourier mode analysis shown in Figure
\ref{mode:a100} looks distinctly different from all previous mode
plots: the $m=2$ mode decays very quickly in parallel with rapidly
growing ${m}=1$ and $3$ modes, which finally surpass the ${m}=2$
mode in amplitude. These distinct features suggest that the
initially ellipsoidal structure of model {\bf A100} was destroyed by
a process that is entirely different from the {\it gradual} decay of
the bar-mode amplitude seen in our models that have been categorized
as stable.  As in model {\bf A134}, the growth time for both the
$m=1$ and 3 modes are very close to each other (as recorded in Table
\ref{table:models}, $T_1= 1.4 T_\mathrm{dyn}$ and $T_3= 1.3
T_\mathrm{dyn}$), but they ultimately saturated at different levels.
We evolved model {\bf A100} for a long time ($\sim 50
T_\mathrm{dyn}$) until the system reached a new, nearly axisymmetric
state.

\begin{figure}[ht]\epsscale{1} \plotone{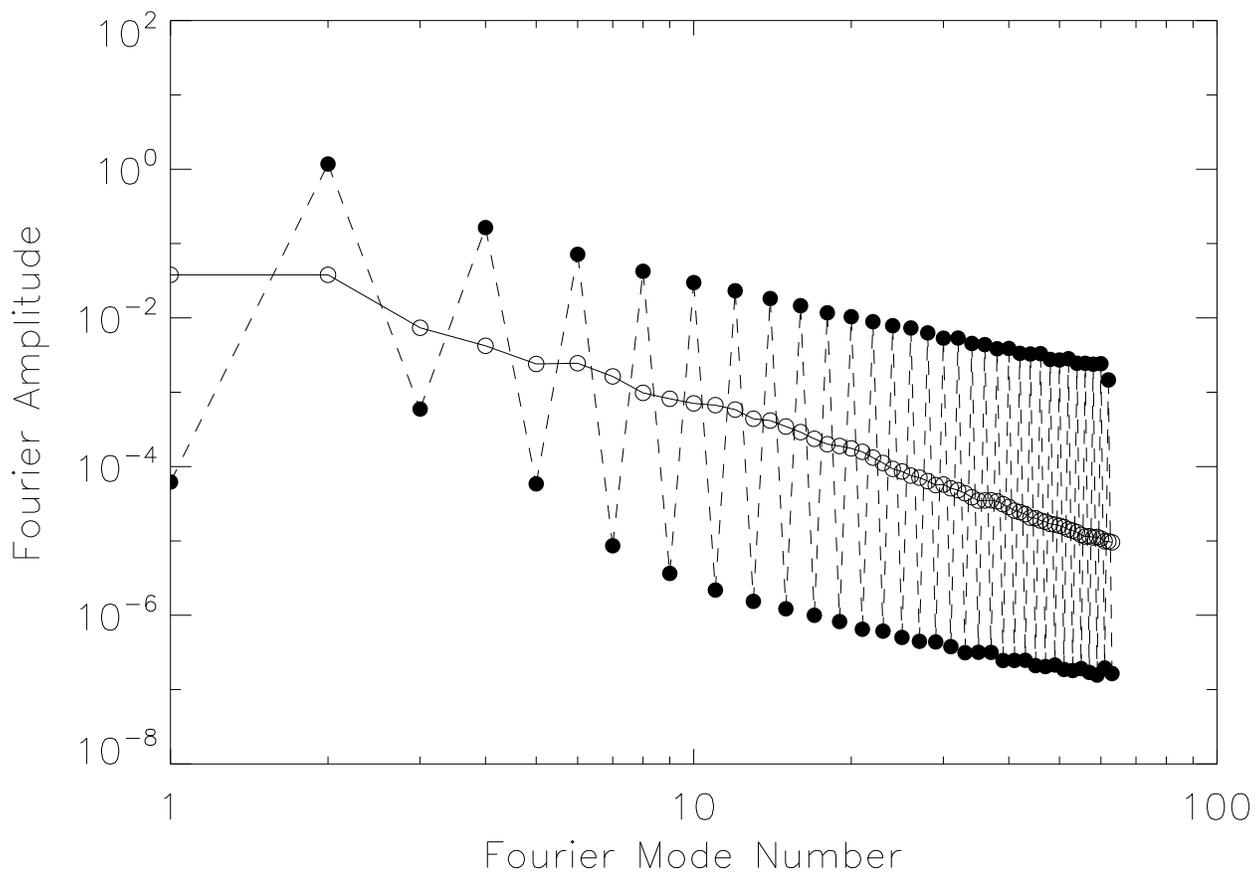}
\caption{
    The azimuthal Fourier mode power spectrum for model {\bf A100}
    initially (filled circles) and at late times (open circles).
    To guide the eye, amplitudes determined for various modes
    at the same time are connected by straight line segments.
    Initially, all the even modes have higher amplitude
    than their neighboring odd modes,
    whereas a cascading power-law distribution appears
    at late times.
    \label{spectrum:a100} }
\end{figure}

The turbulent behavior of model {\bf A100} is very similar to the
outcome of an instability that arose during the late evolution of
the GRR-driven neutron star from a previous study \citep{OTL04}. In
an effort to compare this earlier simulation result with our present
one, we have plotted in Figure \ref{spectrum:a100} a power spectrum
of the various azimuthal modes for model {\bf A100} at time $t=0$
and $t=45 T_\mathrm{dyn}$ in the evolution.
This power spectrum looks surprisingly
similar to the power spectrum that was generated by \cite{OTL04}
late in their model's evolution. The spectrum is dominated by even
modes in the beginning as a result of the ellipsoidal (bar-like)
structure; whereas a turbulence-like power law distribution
dominates at late times, after the energy initially stored in the
$m=2$ bar-mode has cascaded into all other Fourier modes. This
strongly suggests that the violent instability seen in our present
study is the same instability that arose in the \cite{OTL04}
investigation. Because this turbulence-like instability has arisen
in two quite different contexts --- \cite{OTL04} discovered it while
studying the non-linear evolution of an initially axisymmetric star
that was driven to a bar-like configuration by GRR forces, whereas
we have discovered it during purely hydrodynamic evolutions of
compressible triaxial models that have been purposely constructed to
resemble adjoint configurations of Riemann S-type ellipsoids --- it
seems likely that the instability is generic to adjoint
configurations.

\begin{figure}[ht]
\epsscale{0.9} \plotone{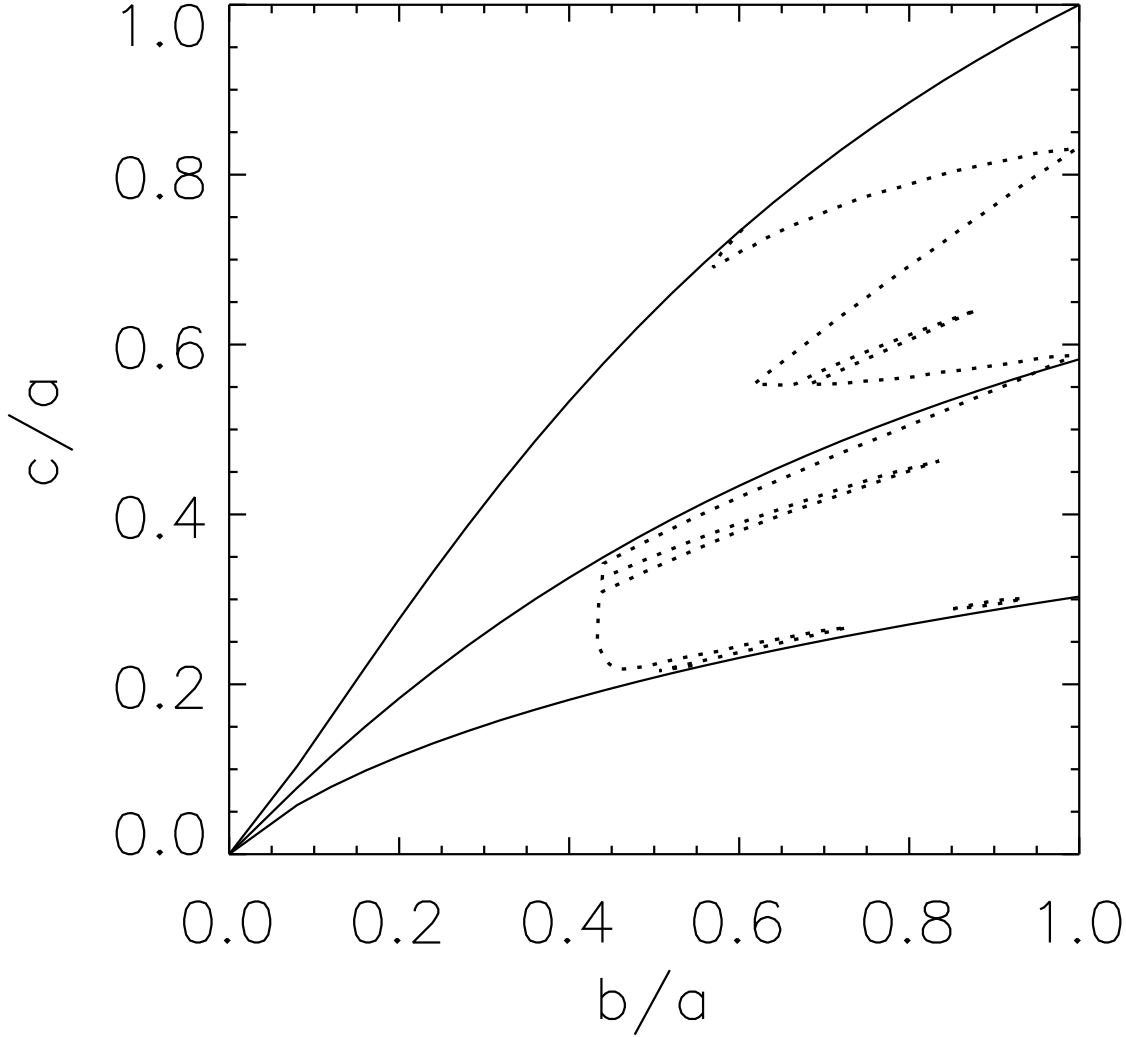} 
\caption{Results from Figure 3a
of \cite{LL96} are redrawn here to provide background to our present
investigation.  Configurations that belong to the adjoint family of
incompressible, Riemann S-type ellipsoids exist in this $(b/a,c/a)$
parameter domain everywhere between the upper solid curve -- drawn
from the origin $(b/a,c/a)=(0,0)$ to $(b/a,c/a)=(1,1)$ -- and the
lower solid curve -- drawn from the origin to $(b/a,c/a)=(1,0.303)$.
Dedekind configurations lie along the solid curve that connects the
origin to $(b/a,c/a)=(1,0.583)$. The jagged, dotted line that
connects the upper curve to the lower curve marks the boundary
between adjoint configurations that are stable (ellipsoids to the
right of the line) and unstable (ellipsoids to the left of the line)
toward the growth of $m=3$ perturbations, as determined from the
linear stability analysis of \cite{LL96}. Note that in certain
regions, a thin ``tongue'' of unstable configurations penetrates
into the region that otherwise contains stable ellipsoids.
    \label{LLplot}}
\end{figure}

\begin{figure}[ht]
\epsscale{0.9}  \plotone{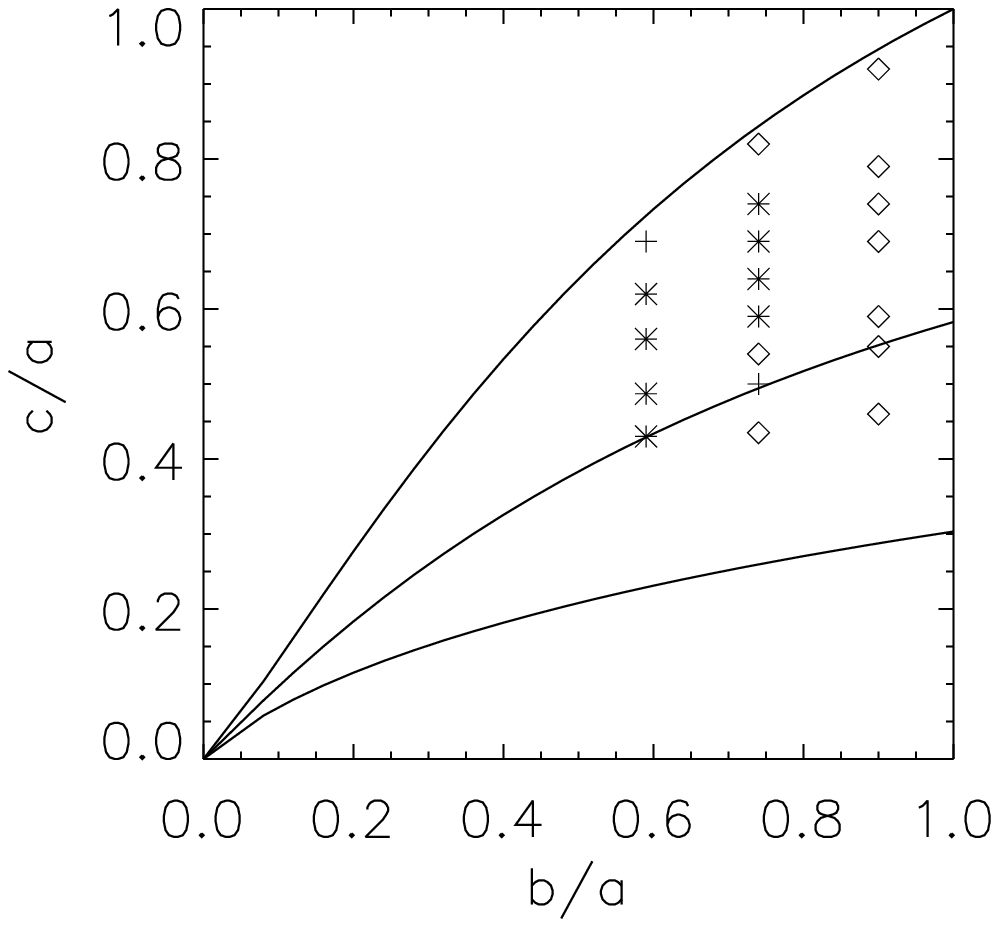} 
\caption{ Results from our
present investigation are summarized across the $(b/a,c/a)$
parameter domain from twenty separate dynamical simulations of
compressible, adjoint configurations. The three solid curves are the
same as in Figure \ref{LLplot}.  Asterisks denote violently unstable
models, plus signs denote moderately unstable models, and diamonds
denote stable models.  The four models whose evolutions have been
discussed in detail in the text can be identified by the $(b/a,c/a)$
parameters defining their initial shapes, as documented in Table 1:
{\bf A010} (0.74,0.821); {\bf A067} (0.90,0.692); {\bf A134}
(0.74,0.487); and {\bf A100} (0.59,0.487). Of the models identified
here, only model {\bf A067} was constructed using the softer, $n=1$
polytropic EOS.
    \label{triaxial}}
\end{figure}

In an effort to understand whether or not this instability occupies
the same regions of parameter space as the elliptical instability
studied by \cite{LL96}, our hydrodynamic investigation has included
models that cover as much parameter space as possible. In Figure
\ref{triaxial}, we have mapped the results from 20 of our adjoint
model evolutions onto the $(b/a, c/a)$ plane. (For the sake of
brevity, a detailed discussion of only 6 of these model evolutions
has been presented, above.) Models found to be violently unstable
are marked by asterisks, stable models are marked by diamonds, and
moderately unstable models are marked by plus signs.  This figure
should be compared with Figure 3a from \cite{LL96} -- reproduced
here as Figure \ref{LLplot} -- which shows the instability domain
occupied by the $m=3$ mode in the adjoint family of incompressible,
Riemann S-type ellipsoids. It is very encouraging that our regions
of instability are consistent with the \cite{LL96} diagram. More
specifically, both investigations find that unstable configurations
occupy a large portion of parameter space above the Dedekind
sequence when $b/a < 0.9$. The growth time also appears to be
related to the axis ratio, $b/a$: models with smaller $b/a$ undergo
a more violent instability. Despite the fact that the method
described in Paper I prevents us from constructing compressible
models with $c/a \lesssim 0.4$ or $b/a \lesssim 0.5$, our results
already indicate that models that are most susceptible to the
instability occupy a smaller fraction of the available parameter
space {\it below} the Dedekind sequence than above it. The two
models that we have categorized as being moderately unstable (the
plus signs in Figure \ref{triaxial}) mark the transition from stable
to violently unstable domains.  In particular, the fact that model
{\bf A134} with $(b/a,c/a)=(0.74,0.487)$ is only moderately unstable
suggests that we may be seeing evidence for the ``tongue'' structure
displayed in Figure \ref{LLplot} that covers most of the Dedekind
sequence. As a counter example, however, we note that all of our
models with $b/a = 0.89$ appear to be stable whereas \cite{LL96}
found that some models with $b/a \sim 0.9$ fall in the instability
domain covered by a set of small extended tongues. This apparent
discrepancy probably arises because the growth rate of the
instability is sufficiently slow in this regime that we are unable
to identify the instability, given the limited evolutionary time of
our simulations.

Because our models are constructed on a rather coarse, discrete
parameter grid, we are unable to map out the instability domain with
the type of fine structure shown in Figure \ref{LLplot}. However,
the fact that the instability domain identified in our present study
and displayed in Figure \ref{triaxial} by and large agrees with the
instability domain identified by \cite{LL96} gives us confidence
that the turbulence-like behavior observed in our violently unstable
models arises from the elliptical instability.

\section{Conclusions and Discussions}

We have conducted a survey of the nonlinear hydrodynamic stability
of compressible triaxial configurations that are initially in
quasi-equilibrium and share the same velocity field as that of
Riemann S-type ellipsoids. Our simulations show that many of these
models are indeed very good long-lived, quasi-equilibrium states. In
addition, we have found that a turbulence-like dynamical instability
arises over a fairly large region of the examined parameter domain
(see Figure \ref{triaxial}). This instability is probably the same
instability observed in the late evolution of a neutron star that
was driven to a bar-like structure by GRR forces \citep{OTL04}. The
characteristics exhibited by this instability and the domain it
occupies in the examined parameter space agree qualitatively with
the elliptical instability that has been discovered in previous
linear studies of incompressible configurations \citep{LL96}.
Therefore, we suspect that this instability is the elliptical
instability that appears to develop generically in fluid flows with
elliptical stream lines.

As a result of this instability, we found that all odd azimuthal
modes grew on a dynamical time scale, but were dominated by the
$m=1$ and $3$ modes, whose amplitudes eventually surpassed the $m=2$
bar-mode and destroyed the ellipsoidal structure of each violently
unstable model. It appears that the growth rate for unstable modes
is higher in models with smaller $b/a$ values. This makes sense
because the elliptical instability ought to disappear in circular
flows with $b/a=1$. In a linear analysis of the elliptical
instability in Riemann S-type ellipsoids, \cite{LL96} did not
discuss the behavior of the $m=1$ mode. However, the appearance of
an eccentric $m=1$ mode in accretion disks \citep{G93,RG94} is
consistent with our discovery here that the $m=1$ mode can also
develop in compressible flows and drain energy from the $m=2$ mode.

The existence of the elliptical instability raises concerns
regarding the final fate of neutron stars that encounter the
GRR-induced secular bar-mode instability. According to previous
theoretical investigations \citep{DL77, LS95}, such a star would be
expected to evolve through a sequence of Riemann S-type ellipsoids
toward the Dedekind sequence that has a vanishingly small pattern
frequency as viewed from the inertial reference frame. However,
\cite{LL96} have shown that most of the Dedekind sequence falls in
the parameter regime where fluid configurations should be
susceptible to the elliptical instability; if any star enters this
region of parameter space, the elliptical instability would set in
on a dynamical time scale. It was unclear in the linear study of
\cite{LL96} whether the elliptical instability would be mild or
violent in the nonlinear regime.  This picture was partially
answered by the nonlinear evolution of one rotating neutron star
that was susceptible to the GRR-induced secular bar-mode instability
\citep{OTL04}; the bar-like configuration was observed to be
destroyed very quickly when the pattern frequency of the bar dropped
to nearly zero. The evolutions of compressible triaxial models
presented in this paper further clarify what the nonlinear outcome
will be of the elliptical instability.  We have found that the
instability can be violent and capable of destroying the ellipsoidal
structure of the star. Therefore, it seems unlikely that a secularly
unstable star driven by GRR forces will actually be able to evolve
along a sequence of Riemann S-type ellipsoids toward a Dedekind-like
configuration.


At the final stage of the evolution reported by \cite{OTL04}, after
the bar-like structure was destroyed and the star returned back to a
nearly axisymmetric state with strong differential rotation, its
$T/|W|$ value was still above the critical limit of $0.14$ for the
secular bar-mode instability. Hence, it is possible that such a star
would again attempt to evolve toward the Dedekind-sequence under the
influence of GRR forces. Cycles of instability could continue until
the star's rotational energy is drained below the critical limit. If
this picture is correct, it would provide a very interesting,
recurrent source of gravitational-wave radiation for ground-based
gravitational-wave detectors such as the Laser Interferometer
Gravitational-wave Observatory (LIGO) and its sister instruments
worldwide.

\acknowledgments

We would like to thank Lee Lindblom, John Friedman, Nikolaos Stergioulas,
Yuk Tung Liu and Juhan Frank for valuable discussions.
This work was partially supported by NSF grants AST-0407070 and PHY-0326311.
The computations were performed on the Supermike cluster and
the Superhelix cluster at LSU, and on the Tungsten cluster
at National Center for Supercomputing Applications (NCSA).

\end{document}